\documentclass[aps,prb,twocolumn,amsmath,amssymb,floatfix,superscriptaddress,longbibliography]{revtex4-2}
\usepackage{natbib}
\usepackage{graphicx,xcolor,multirow,bm,hyperref}
\usepackage{float}
\usepackage[ignoreunlbld,norefs,nocites]{refcheck}
\begin{document}

\title{Electronic Temperature–Driven Phase Stability and Structural Evolution of Iron at High Pressure}
\author{S. Azadi}
\affiliation{Department of Physics and Astronomy, University of Manchester, Oxford Road, Manchester M13 9PL, United Kingdom}
\affiliation{Department of Physics, Clarendon Laboratory, University of Oxford, Parks Road, Oxford OX1 3PU, United Kingdom}
\email{sam.azadi@physics.ox.ac.uk}
\author{S.M. Vinko}
\affiliation{Department of Physics, Clarendon Laboratory, University of Oxford, Parks Road, Oxford OX1 3PU, United Kingdom}

\begin{abstract}
We present Gibbs free-energy phase diagrams for compressed iron within a pressure range of 20 to 300 GPa and electronic temperature up to 3 eV obtained using finite-temperature density functional and density functional perturbation theories. Our results for bcc, fcc, and hcp phases predict solid-solid phase transitions in iron driven purely by electronic entropy and temperature. We found a phase transition from hcp to bcc at pressures above 200 GPa, which depends on the electronic temperature. An experimental observation of the stability of the bcc phase above 200 GPa by X-ray Free Electron Laser has recently been reported \cite{Konopkova}. 
\end{abstract}
\maketitle

\section{Introduction}
Iron, one of the most abundant elements in the Earth’s interior and a cornerstone of modern technology, exhibits a rich and complex phase diagram when subjected to extreme conditions of pressure and temperature\cite{Pepperhoff,Anzellini,Belonoshko,Vocadlo,Wilson}. Under ambient conditions, iron crystallizes in the body-centered cubic (bcc) structure, known as $\alpha$-iron or ferrite. However, as pressure and temperature increase, such as in planetary cores or during high-energy experiments, iron undergoes a series of structural and electronic transitions, including transformations to face-centered cubic (fcc), hexagonal closed-packed (hcp), and possibly bcc phases \cite{Ping2013,Dubrovinsky,Hwang2021}. Understanding this behavior is essential for fields ranging from geophysics and planetary science to inertial confinement fusion and high-energy-density physics.

Historically, the investigation of the iron phase diagram under extreme conditions has focused primarily on its ionic temperature, that is, the temperature associated with the motion of the atomic nuclei. Experimental techniques such as diamond anvil cells (DAC) and dynamic compression (e.g., shock or ramp loading) have been extensively used to probe iron structural transitions by measuring pressure and lattice temperature \cite{Dewaele2006,Merkel2006,Shen1998}. These approaches have yielded a reasonably detailed map of the solid-solid and solid-liquid boundaries in the pressure-temperature (P-T) space, particularly at conditions relevant to the Earth’s core.

In contrast, less attention has been paid to the role of electronic temperature, which describes the thermal excitation of electrons independently of the lattice. Under conditions of ultrafast laser heating or strong compression, electrons can become significantly hotter than the ions, resulting in a non-equilibrium state known as a two-temperature system \cite{Alexopoulou2024}. Such electronic excitations can alter the free energy landscape, destabilize equilibrium phases, and even induce novel transient states that are inaccessible under equilibrium conditions. The limited exploration of these effects represents a significant gap in our understanding, particularly as electronic temperature plays a pivotal role in governing thermodynamic and transport properties in high-energy-density environments \cite{Siders,Sundaram,Recoules,Ernstorfer}.

Therefore, a more comprehensive understanding of the behavior of different phases of iron under extreme conditions requires bridging this gap by incorporating both ionic and electronic contributions to the free-energy landscape. This dual perspective is increasingly relevant not only for modeling planetary interiors but also for interpreting time-resolved experiments using X-ray free-electron lasers (XFELs) and high-intensity optical pulses, which can drive matter far from equilibrium\cite{Rousse,Widmann,Sciaini,Collet,Amouretti2025,Celin2025}.

Density Functional Theory (DFT) is a suitable and widely adopted first-principles method to calculate the phase diagram of iron under extreme conditions \cite{Stixrude1995,Neumann2001,Vocadlo2012,Belonoshko2000}. Its ability to accurately capture the quantum-mechanical behavior of electrons makes it a useful tool for modeling structural, electronic, and magnetic properties across a broad range of pressures and temperatures. This is particularly important under extreme conditions, where experimental data are limited or difficult to obtain and empirical interatomic potentials may fail to extrapolate reliably. Within DFT and electronic structure frame work, the entropic pressure refers to the contribution of entropy to the effective pressure exerted by the electron density distribution. This concept arises primarily in systems where thermal or configurational entropy plays a significant role, such as warm dense matter, high-temperature physics, plasma physics, and DFT calculations at finite temperatures. In temperature-dependent DFT (Mermin-DFT) \cite{Mermin}, fractional occupation numbers of orbitals introduce an entropic term in the free-energy functional, leading to entropic contributions to the potential and pressure.

In the context of finite-temperature DFT, the Helmholtz free energy F is given by $F = E - TS$, where $E$ is the total electronic energy (kinetic + potential), $T$ is the temperature and S is the electronic entropy, typically written as $S = - k_B \sum_i [ f_i \ln f_i + (1 - f_i) \ln (1 - f_i) ]$, where $f_i$ are the occupation numbers of electronic states. Taking the derivative of F with respect to volume yields an effective entropic pressure term $P = -\left(\frac{\partial F}{\partial V} \right)_{T,N}$, which includes an entropic contribution due to the occupancy distribution. 

If the volume is fixed (constant density), the traditional thermodynamic definition of pressure $P = -\left(\frac{\partial F}{\partial V} \right)_{T,N}$ does not apply directly because there is no volume change. However, in electronic structure theory, particularly at finite temperatures, other definitions of pressure can still be relevant. Even when the volume is fixed, one can define an internal pressure-like quantity using the stress tensor $P = -\frac{1}{3} \sum_{\alpha} \sigma_{\alpha\alpha}$ where $\sigma_{\alpha\alpha}$ with $\alpha=x,y,z$  are the diagonal components of the stress tensor $\sigma$, which is calculated as $\sigma_{\alpha\beta} = \frac{1}{V} \left( \sum_i m_i v_{i\alpha} v_{i\beta} - \sum_{i<j} r_{ij\alpha} F_{ij\beta} \right)$ where $m_i$ and $v_i$ are the mass and velocity of the particle $i$, $r_{ij}$ is the distance between the particles $i$ and $j$, $F_{ij}$ is the force between them. In DFT calculations, the stress tensor includes both the kinetic and exchange-correlation contributions. For a fixed volume confined system, the contribution of entropy to internal stress can be derived from the Helmholtz free energy F: $\sigma_{\alpha\beta}^{\text{entropy}} = -\frac{1}{V} \left( \frac{\partial (TS)}{\partial \epsilon_{\alpha\beta}} \right)_{T,N}$, where $\epsilon_{\alpha\beta}$ is the strain tensor. This term accounts for the entropy-driven contribution to the internal forces. The entropic pressure-like quantity in a fixed-volume system comes from the entropy term in the stress tensor: $P_{\text{entropy}} = - \frac{1}{3} \sum_{\alpha} \left( \frac{\partial (TS)}{\partial \epsilon_{\alpha\alpha}} \right)_{T,N}$ This measures how entropy variations modify internal stresses. For Fermi-Dirac electrons in a highly excited system, the entropy per unit volume is: $S = -k_B \int g(\epsilon) \left[ f \ln f + (1 - f) \ln (1 - f) \right] d\epsilon$ where $g(\epsilon)$ is the electronic density of states, and $f = (1 + e^{(\epsilon - \mu)/k_B T})^{-1}$ is the Fermi-Dirac distribution. Thus,the entropy-driven internal stress can be approximated as: $\sigma_{\alpha\alpha}^{\text{entropy}} = - k_B T \int g(\epsilon) \frac{\partial f}{\partial \epsilon} d\epsilon$ which acts as an internal pressure-like force.

The entropic pressure in a confined system can induce a solid-solid phase transition\cite{Azadi2024,Azadi2025}. The fundamental reason is that entropy affects free energy, which in turn influences the stability and phase behavior of the system. Even if the volume is fixed, entropy-driven internal stresses can alter the relative stability of different solid phases, leading to a phase transition. Solid-solid phase transitions in warm dense matter (WDM) or other high-energy-density conditions occur.  Since entropic pressure contributes to the total stress tensor $\sigma_{\alpha\beta}$, it can modify the landscape of free energy and drive a structural transition. A solid-solid transition occurs when two different crystalline structures have the same free energy at some critical temperature or stress condition. The stability of a phase is determined by the second derivative of $F$: $\left(\frac{\partial^2 F}{\partial \lambda^2} \right)_{T, V} > 0$, where $\lambda$ is a structural parameter (e.g., lattice distortion, atomic displacement). If entropy effects modify the free energy so that this stability criterion changes, a phase transition can occur.

In this work, by calculating the Gibbs free energy difference $\Delta G = \Delta H - T\Delta S$ between the iron phases, where G, H, T, and S are the Gibbs free energy, enthalpy, electronic temperature, and entropy, respectively, we predict the stability of the fcc and bcc phases compared with hcp driven by electronic entropy.

\section{Calculation details}
To calculate the thermodynamic phase diagram of iron, we use the Quantum Espresso package~\cite{QE}, with the Perdew-Burke-Ernzerhof (PBE) parameterization~\cite{PBE} of the exchange-correlation (XC) functional. We considered bcc, fcc, and hcp crystal structures of Fe at electronic temperatures of up to 3 eV and pressure of up to 300 GPa, using a kinetic-energy cutoff of the plane-wave basis set of 80~Ry and an augmentation charge energy cutoff of 960 Ry. The finite-temperature DFT and density functional perturbation theory (DFPT) \cite{Baroni} calculations were performed using a $(24,24,24)$ $\mathbf{k}$-grid and a $(8,8,8)$ $\mathbf{q}$-grid, respectively. The ultrasoft tabulated pseudopotential~\cite{QE2} with 16 valence electrons was used. The electronic temperature is controlled by the standard Fermi-Dirac distribution function. The number of empty bands increases with the electronic temperature. The temperature parameter in this work is related only to electrons as the temperature of the ions is fixed at zero. The purpose of this assumption is to purely study the effect of electronic entropy and temperature on the phase diagram of iron.

\section{Results and discussion}
The Gibbs free can be expressed as $G = G_{el} + G_{lat}$ where $G_{el}$ is the electronic Gibbs free energy without lattice dynamics and $G_{lat}$ is the phonon Gibbs free energy. We calculated $G_{el}$ for the phases bcc, fcc and hcp up to 300 GPa and the electronic temperature up to 3 eV. To span the $G_{el}(P,T)$ phase diagram, we carried out 800 separate DFT calculations for each phase under different P, T conditions. The geometry of the systems at each DFT calculation was fully optimized. We interpolated the energy difference between two phases on a $10^3\times 10^3$ grid.

\begin{figure}
    \centering
      \includegraphics[width=1.\linewidth]{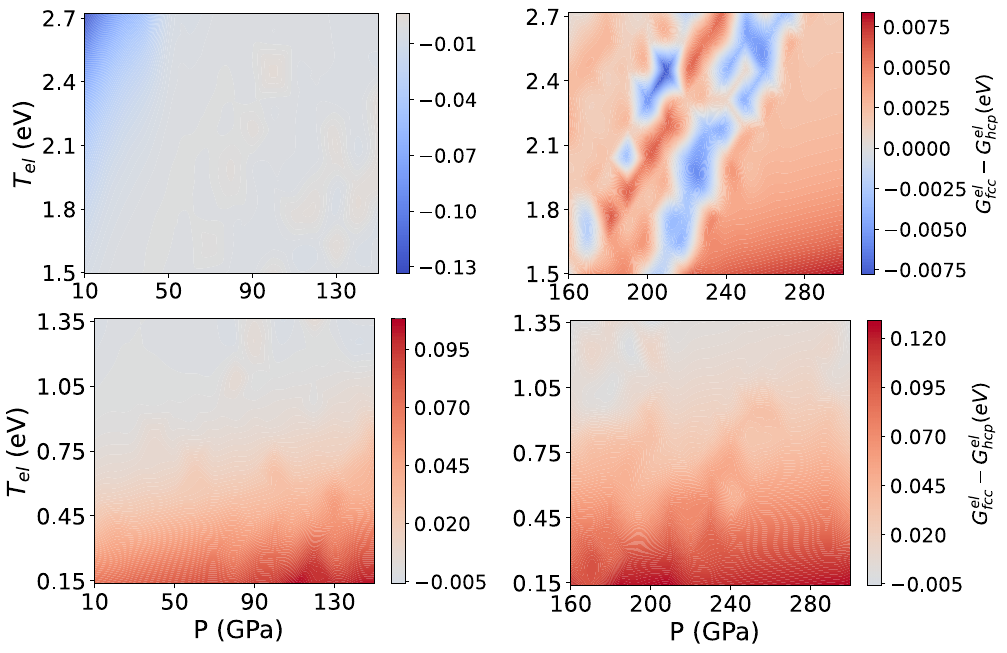}   \\
    \caption{The Gibbs free energy difference between fcc and hcp phases. 1600 finite-temperature DFT simulations were performed (800 for each phase), where the lattice parameters of the hcp phase (both a and c) and the fcc phase were fully optimised to satisfy the input T$_{el}$ and P conditions.}
    \label{fig:fcc2hcp}
\end{figure}

Figure~\ref{fig:fcc2hcp} shows the contour plot energy difference between hcp and fcc phases. The effect of electronic temperature on the stability of the fcc phase can be clearly observed. The Gibbs free energy difference between the fcc and hcp phases reveals a nuanced interplay between pressure and electronic temperature. At electronic temperatures below 1.3 eV, increasing pressure stabilizes the hcp phase relative to fcc. However, as the electronic temperature rises, which can occur in laser-heated or shock-compressed environments, the fcc phase gains thermodynamic favorability because of its higher electronic entropy. This results in a competition between the pressure-driven hcp stability and the temperature-driven fcc stabilization.

At pressures and temperatures above 150 GPa and 1.4 eV, respectively, this interplay produces a complex phase stability landscape, where regions of fcc and hcp dominance emerge with overlapping phase boundaries. The Gibbs free energy surfaces of the two phases approach near-degeneracy in certain P–T regimes, leading to mixed-phase domains and potential transient coexistence during rapid dynamical processes. These results underscore the importance of accounting for both the ionic and electronic contributions to free energy in modeling the phase behavior of iron under extreme conditions.

\begin{figure}
\centering
\includegraphics[width=0.75\linewidth]{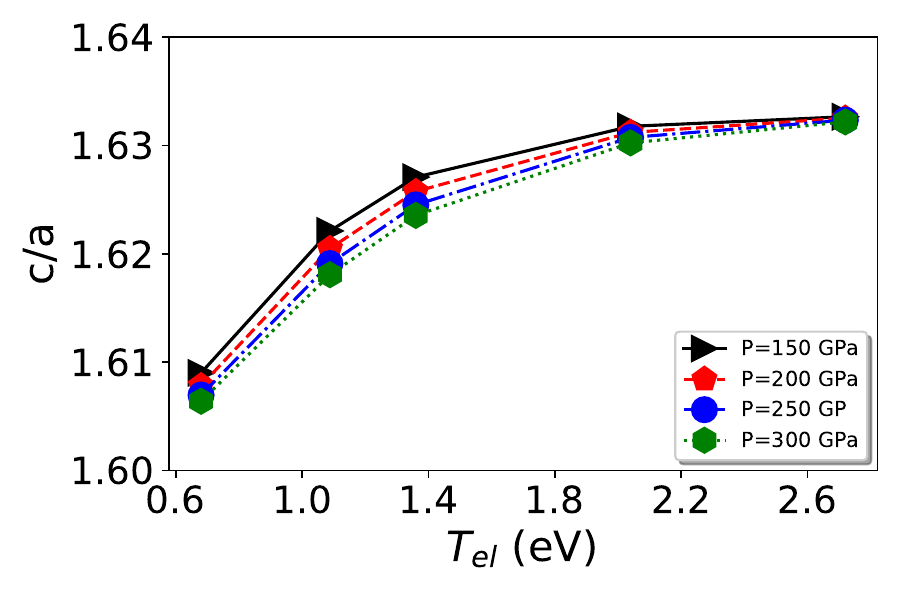}\\
\includegraphics[width=0.75\linewidth]{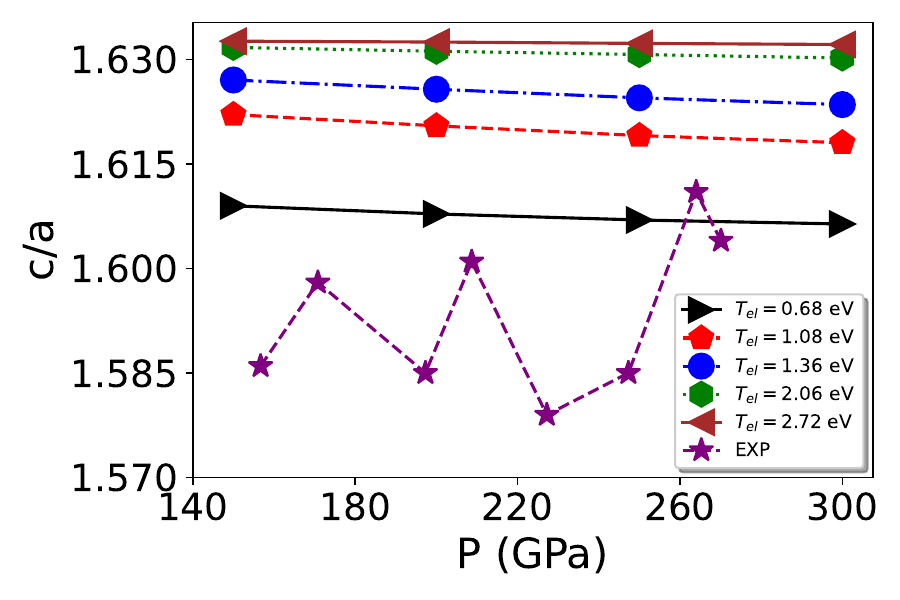} 
\caption{\label{fig:hcp_c_a} (Left panel) Evolution of $c/a$ parameter of hcp phase as a function of pressure obtained at different electronic temperature T$_{el}$. The experimental data are obtained using X ray diffraction experiments with the diamond anvil cell\cite{Mao1990}. (Right panel) Non-linear increasing of $c/a$ as a function of electronic temperature.}
\end{figure}

We computed the optimized axial ratio $c/a$ of the iron hcp phase as a function of the pressure and electronic temperature illustrated in Fig.\ref{fig:hcp_c_a}. At fixed pressure, we find that increasing the electronic temperature leads to a systematic increase in the c/a ratio. However, this increase is markedly nonlinear as it follows an exponential trend with Pad\'{e} asymptote at high temperatures. For example, at 150 GPa, raising the electronic temperature from 0.68 eV to 1.09 eV results in an increase in c/a from 1.609 to 1.623. In contrast, a temperature increase from 2.0 to 2.7 eV yields a much smaller change, from 1.6317 to 1.6326, highlighting the saturation behavior.

This nonlinear dependence of the c/a ratio on the electronic temperature appears to be largely independent of pressure and offers insight into the thermal destabilization of the hcp phase. As the c/a ratio increases with temperature, the hcp lattice becomes progressively distorted away from its equilibrium configuration, which may contribute to the softening of specific phonon modes and the eventual instability of the hcp structure at elevated electronic temperatures. This behavior is consistent with the observed preference for the fcc phase under high-temperature, high-pressure conditions, where electronic excitations play a dominant role.

In contrast to the effect of electronic temperature, increasing pressure leads to a slight but linear decrease in the optimized $c/a$ ratio of the hcp phase (Fig.\ref{fig:hcp_c_a}). This behavior, which is independent of the electronic temperature, highlights a fundamental opposition in how pressure and electronic temperature influence the lattice geometry. While electronic excitation expands the c-axis relative to the a-axis, driving the $c/a$ ratio upward, increased pressure tends to compress the structure more uniformly, reducing the $c/a$. This opposing trend further illustrates the competing roles of pressure and electronic temperature in determining the structural stability of the hcp phase under extreme conditions.

\begin{figure}
\centering
\includegraphics[width=0.95\linewidth]{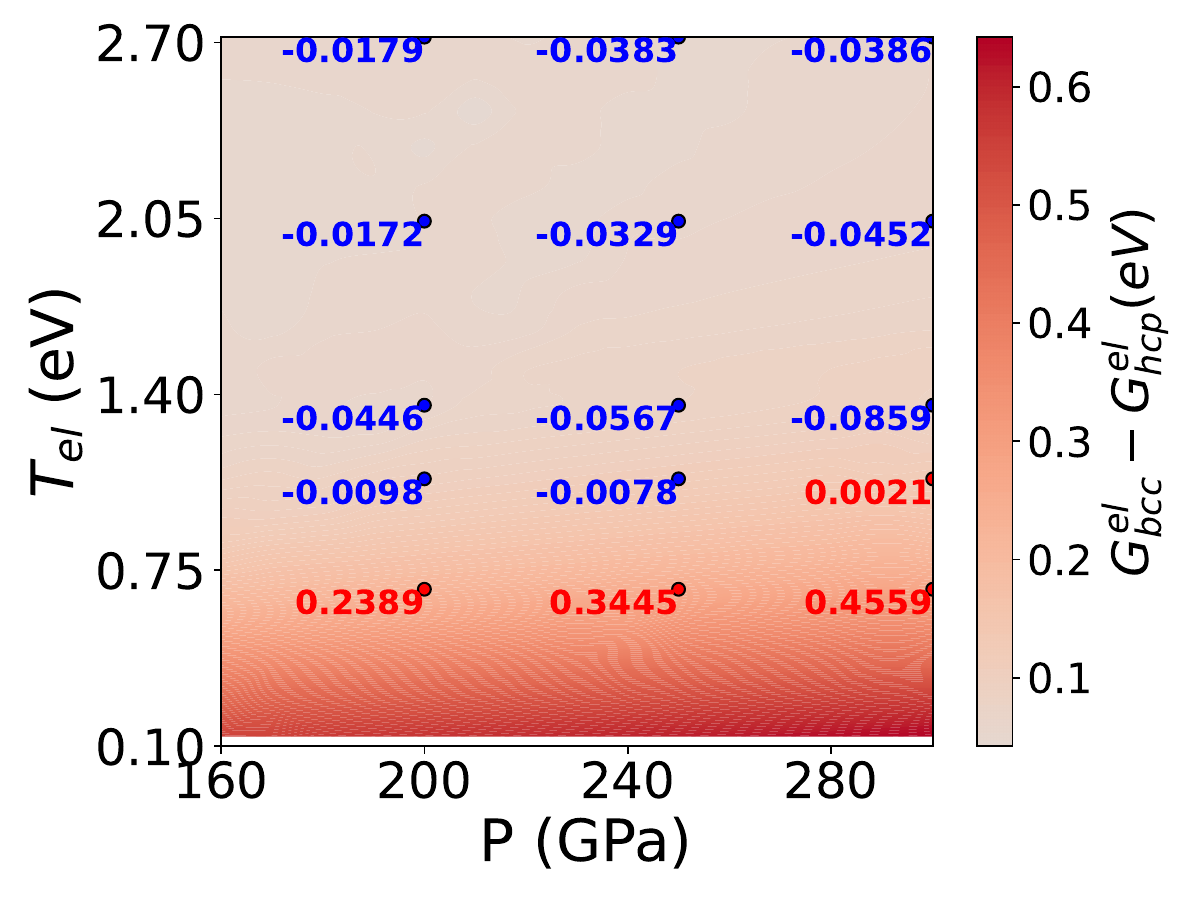}
\caption{\label{fig:hcp2bcc} The Gibbs free energy difference between hcp and bcc phases. The color map shows the electronic Gibbs free energy difference. The data points on the map show the full (electronic and lattice dynamic) Gibbs free energy difference.}
\end{figure}

We computed the electronic Gibbs free energy difference between the hcp and bcc phases while initially neglecting phonon effects (Fig.~\ref{fig:hcp2bcc}) . Within the pressure and temperature range considered, above 150 GPa and electronic temperatures exceeding 0.1 eV, the results show that there is no region in the phase space where the bcc is energetically favored over the hcp phase. In this approximation, the hcp structure remains consistently more stable, which is opposite to the fcc-hcp phase diagram in Fig.~\ref{fig:fcc2hcp}.

However, when the contribution of the phonon to the Gibbs free energy is included for both phases, the stability landscape changes significantly, as highlighted by the data points with their values in Fig.~\ref{fig:hcp2bcc}. The total Gibbs free energy difference reveals that the bcc phase becomes thermodynamically stable over the hcp phase as the electronic temperature increases. This highlights the crucial role of lattice vibrations in stabilizing the bcc structure under high-temperature conditions. The phonon entropy of the bcc phase, which is generally higher because of its greater structural symmetry and softer vibrational modes, contributes significantly to lowering its free energy at elevated temperatures. These findings underscore the importance of including both electronic and vibrational contributions when assessing phase stability under extreme conditions.

\begin{figure}
\centering
\includegraphics[width=0.95\linewidth]{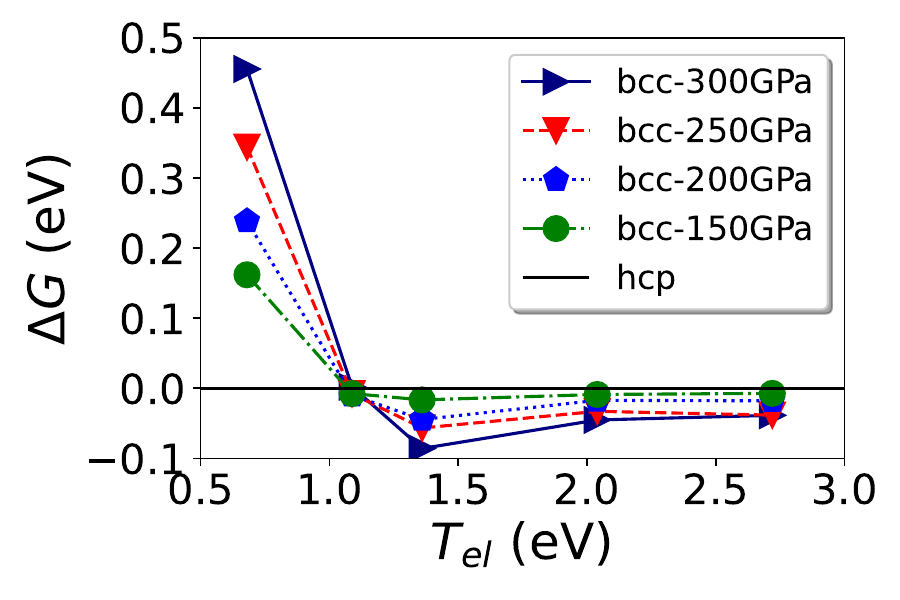}
\caption{\label{fig:phasediagram} Thermodynamic Gibbs free energy of Fe-bcc with respect to Fe-hcp as a function of electronic temperature. Free energies are calculated at pressures 150, 200, 250, and 300 GPa.}
\end{figure}

The Gibbs free energy phase diagram of the bcc and hcp phases at a pressure range of 150-300 GPa, plotted as a function of electronic temperature in Fig.\ref{fig:phasediagram}, reveals that the free energy difference between the two phases sharply drops with increasing temperature for $T_{el}<1.1$ eV. At electronic temperatures above 1.1 eV, the Gibbs free energies of the bcc and hcp phases converge, indicating near-degeneracy. In this regime, the system approaches a thermodynamic condition where neither phase is distinctly favored, suggesting the possibility of phase coexistence or metastability. This near-equality in Gibbs free energy implies that slight perturbations, such as strain, defects, or dynamic fluctuations, could tip the balance between the two phases, enabling transitions or coexistence on experimental timescales. Such a scenario is particularly relevant in ultrafast or shock-driven processes, where non-equilibrium conditions can stabilize metastable states over extended durations.

\begin{figure*}
    \centering
    \includegraphics[width=0.95\linewidth]{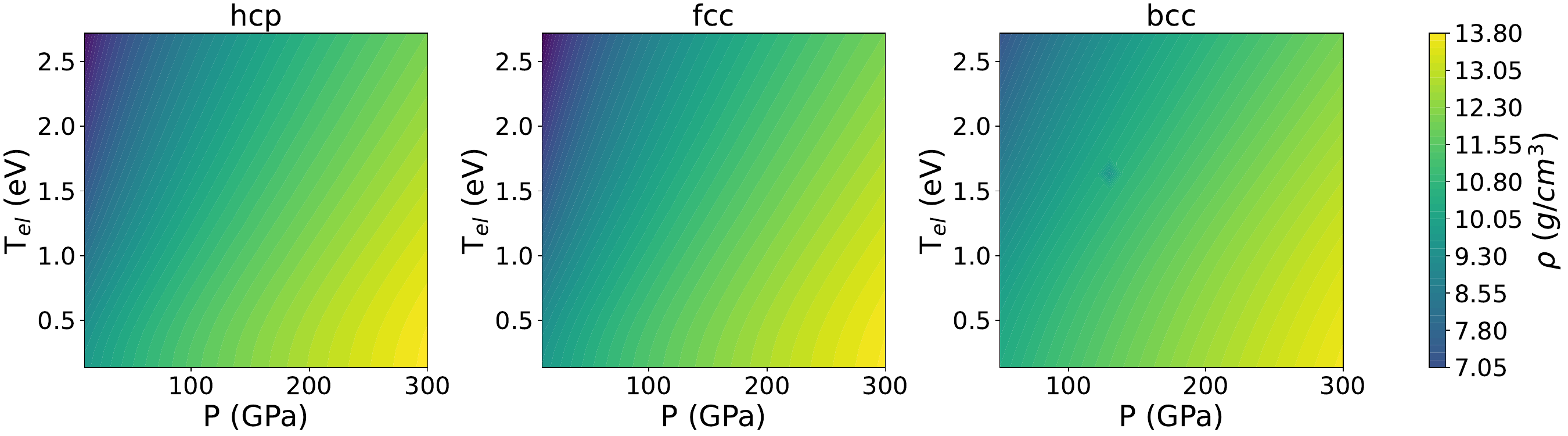}\\
    \includegraphics[width=0.95\linewidth]{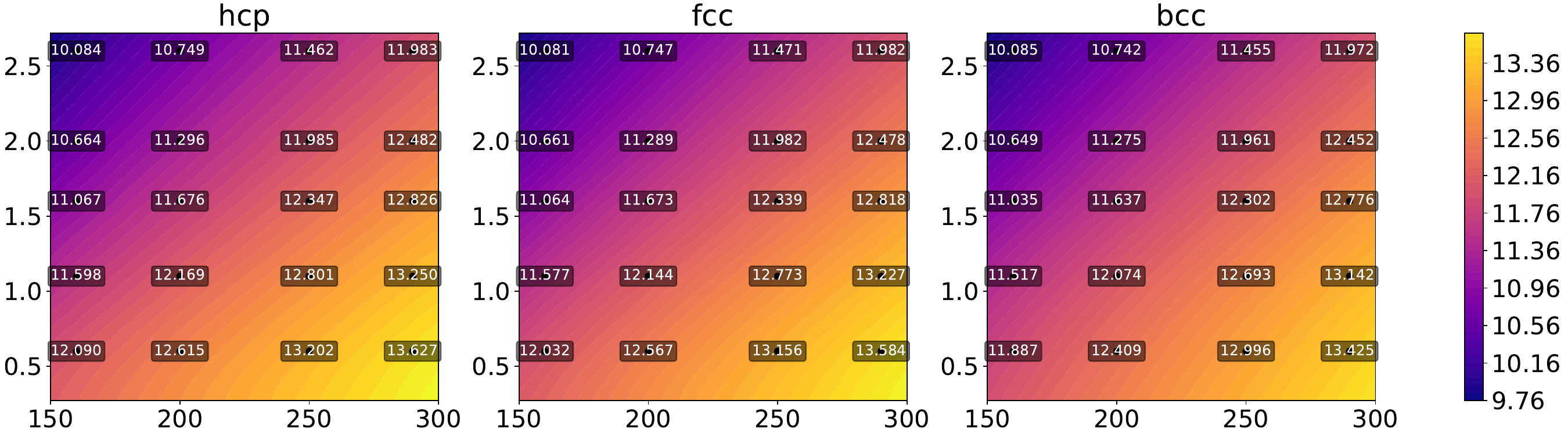}
    \caption{Density of hcp, fcc, and bcc phases of iron in $g/cm^3$ as a function of pressure and electronic temperature. The above panel show the whole studied temperature and pressure range, while the second panel only focuses on high-pressure range. The axes in all plots are the same.}
    \label{fig:density}
\end{figure*}

As the electronic temperature increases, the bcc phase becomes more stable relative to the hcp phase. This trend is reflected in the Gibbs free energy difference, which increases with temperature, indicating a growing thermodynamic preference for the bcc structure. However, this stabilization is not linear with respect to temperature. Instead, the energy difference between the hcp and bcc phases exhibits a maximum around an electronic temperature of $\sim$ 1.4 eV, whose value depends on pressure. This suggests the existence of a distinct temperature at which the bcc phase reaches peak thermodynamic stability relative to hcp. Beyond this point, the energy gain from further increasing the electronic temperature diminishes, indicating that bcc stabilization saturates or may even begin to weaken. This non-monotonic behavior points to a complex interplay of electronic entropy and structural energetics, and it may correspond to a critical regime of metastability or optimal coexistence relevant for ultrafast laser or shock-driven experiments.

A contour plot of the density of the hcp, fcc, and bcc phases of iron as functions of pressure and electronic temperature reveals a clear and systematic trend: increasing pressure leads to higher density across all phases, as expected from compression effects (Fig~\ref{fig:density}) . In contrast, increasing electronic temperature has the opposite effect, reducing the density as the temperature rises. This density reduction arises from electronic excitation, which tends to expand the lattice due to the increased electronic pressure and weakened bonding. The competition between pressure-driven compression and temperature-induced expansion leads to characteristic contour patterns in the phase space. In particular, this inverse relationship between pressure and electronic temperature with respect to density is consistent across the hcp, fcc, and bcc structures, although the magnitude of the effect varies depending on the specific phase and its compressibility.

At low electronic temperatures (below ~1 eV), the density of the hcp phase is higher than that of the fcc and bcc phases across the entire pressure range considered. Specifically, the ordering of the densities follows the relation $\rho_{\mathrm{hcp}} > \rho_{\mathrm{fcc}} > \rho_{\mathrm{bcc}}$. This systematic trend reflects the more efficient atomic packing and a lower specific volume of the hcp structure under these conditions. The higher density of the hcp phase contributes to its thermodynamic stability at low temperatures, as denser phases are typically favored under high pressure when electronic excitations are minimal. This density-driven stabilization likely explains the dominance of the hcp phase across a wide pressure range at low electronic temperatures.

In general, the reduction in density with increasing electronic temperature at fixed pressure is more pronounced in the hcp phase than in the bcc phase. This indicates a higher thermal expansivity of the hcp structure under electronic excitation. For example, at 200 GPa, raising the electronic temperature from 0.6 eV to 2.6 eV decreases the density of the hcp phase by approximately 1.866 $g/cm^3$, whereas the corresponding decrease in the bcc phase is about 1.667 $g/cm^3$. This greater sensitivity of the hcp phase to electronic temperature further supports its relative instability at elevated excitation levels and may contribute to the observed stabilization of the bcc phase under such conditions.

\section {conclusion}
Using first-principles density functional theory, we investigated the structural and thermodynamic behavior of the hcp, fcc and bcc phases of iron under extreme conditions of high pressure and elevated electronic temperature. Our results reveal a complex interplay between pressure and electronic temperature in shaping phase stability, structural parameters, and density.

We find that pressure stabilizes the hcp phase and increases the material density, while electronic temperature has the opposite effect, reducing density and favoring higher entropy phases like bcc and fcc. The c/a ratio of the hcp phase increases non-linearly with the electronic temperature, saturating at high temperatures and slightly decreasing with increasing pressure. This opposing behavior indicates that pressure and temperature influence the lattice geometry in fundamentally different ways.

Gibbs free-energy calculations show that, when only electronic contributions are considered, the hcp phase remains more stable than the bcc phase across the studied pressure-temperature range. However, including phonon contributions reveals that the bcc phase becomes increasingly stable at higher electronic temperatures. Notably, the Gibbs free energy difference between the bcc and hcp phases reaches a maximum around 1.4 eV, suggesting a temperature at which bcc stabilization is strongest. At temperatures above 2 eV, the free energy difference between the phases hcp and bcc becomes very small, implying the possibility of phase coexistence or metastability.

Density trends further support these observations: at low electronic temperatures (T $<$ 1 eV), the ordering $\rho_{\mathrm{hcp}} > \rho_{\mathrm{fcc}} > \rho_{\mathrm{bcc}}$ holds throughout the pressure range, which correlates with the stability of the hcp phase in this regime. As the temperature increases, the density reduction is more significant in the hcp phase than in the bcc phase, indicating a greater thermal expansivity and contributing to the destabilization of hcp in favor of bcc at high temperatures.

Our study highlights the importance of accounting for both electronic and vibrational contributions to free energy when evaluating phase stability under extreme conditions. The non-monotonic and competing effects of pressure and electronic temperature shape a rich and nuanced phase landscape for iron, relevant for understanding its behavior in planetary interiors and high-energy-density environments.

\section{Acknowledgment} 
The authors acknowledge support from the UK EPSRC grant EP/W010097/1 and from the Royal Society. 

\bibliography{main}

\end{document}